\documentclass[prb,twocolumn,preprintnumbers,amsmath,amssymb]{revtex4}

\usepackage{graphicx}

\usepackage{bm}
\usepackage{amsmath}
\usepackage{amssymb}
\usepackage[latin1]{inputenc}
\usepackage{color}
\usepackage[colorlinks=true, pdfstartview=FitV, linkcolor=blue, citecolor=blue, urlcolor=blue]{hyperref} 

\begin{document}

\title{Local dynamics of topological magnetic defects in the itinerant helimagnet FeGe}

\author{A. Dussaux,$^{1,\dagger}$ P. Schoenherr,$^{2,\dagger}$ K. Koumpouras,$^3$ J. Chico,$^3$ K. Chang,$^1$ L. Lorenzelli,$^1$ N. Kanazawa,$^4$ Y. Tokura,$^{4,5}$ M. Garst,$^6$ A. Bergman,$^3$ C. L. Degen,$^{1,\ast}$ and D. Meier$^{2,7\ast}$} 

\affiliation{$^1$ Department of Physics, ETH Zürich, Otto Stern Weg 1, 8093 Zurich, Switzerland}
\affiliation{$^2$ Department of Materials, ETH Zürich, Vladimir-Prelog-Weg 4, 8093 Zurich, Switzerland}
\affiliation{$^3$ Department of Physics and Astronomy, Uppsala University, PO Box 516, 75120 Uppsala, Sweden}
\affiliation{$^4$ Department of Applied Physics, University of Tokyo, Tokyo 113-8656, Japan}
\affiliation{$^5$ RIKEN Center for Emergent Matter Science (CEMS), Wako 351-0198, Japan}
\affiliation{$^6$ Institute for Theoretical Physics, Universität zu Köln, D-50937 Köln, Germany}
\affiliation{$^7$ Department of Materials Science and Engineering, Norwegian University of Science and Technology, 7491 Trondheim, Norway}

\begin{abstract}
$^{\dagger}$ These authors contributed equally to this work\\
$^{\ast}$ degenc@ethz.ch; dennis.meier@ntnu.no\\
\\
Chiral magnetic interactions induce complex spin textures including helical and conical spin waves, as well as particle-like objects such as magnetic skyrmions and merons. These spin textures are the basis for innovative device paradigms and give rise to exotic topological phenomena, thus being of interest for both applied and fundamental sciences. Present key questions address the dynamics of the spin system and emergent topological defects. Here we analyze the micromagnetic dynamics in the helimagnetic phase of FeGe. By combining magnetic force microscopy, single-spin magnetometry, and Landau-Lifschitz-Gilbert simulations we show that the nanoscale dynamics are governed by the depinning and subsequent motion of magnetic edge dislocations. The motion of these topologically stable objects triggers perturbations that can propagate over mesoscopic length scales. The observation of stochastic instabilities in the micromagnetic structure provides new insight to the spatio-temporal dynamics of itinerant helimagnets and topological defects, and discloses novel challenges regarding their technological usage. 
\end{abstract}

\maketitle

\newpage

Intriguing states of magnetism\cite{fert13} arise in transition-metal silicides and germanides of the B20-type such as MnSi\cite{muhlbauer09, ishikawa76}, Fe$_{1-x}$Co$_x$Si\cite{beille81, munzer10, milde13}, and FeGe\cite{lundgren70,lebech89,yu10}. The competition of ferromagnetic exchange, Dzyaloshinskii-Moriya (DM) interaction, and magnetic anisotropy leads to a variety of complex magnetic phases with spins forming helical or conical spirals, as well as long-range ordered lattices of magnetic whirls\cite{rossler06}. These spin structures are appealing as they give rise to anomalous transport properties\cite{neubauer09,schulz12}, exotic vortex domain walls\cite{li12}, and unusual dynamic spin-wave phenomena\cite{janoschek10, koralek12, kugler15}. Of particular interest is the emergence of topologically protected spin states, i.e., stable magnetic configurations that cannot be generated or destroyed by a continuous transformation of the spin system\cite{mermin79}. These topological defects are explicitly robust and may serve as functional objects in future spintronics devices\cite{Tomasello14,Kruglyak10}. At present, however, we are only at the verge of grasping the technological potential of topological spin states \cite{bode07, togawa12} and their complex nanoscale physics is still largely unexplored.

During the last years, research activities in the field mainly focused on topologically protected magnetic whirls called skyrmions. Skyrmions arise in various B20 materials under magnetic fields and represent particle-like entities that can be moved\cite{schulz12}, written, and erased on demand\cite{romming13}. Although it is known that the formation of skyrmions is facilitated by superior topological defects that develop in the helimagnetic ground state\cite{yu_12}, little attention has been paid to the latter ones. In the helimagnetic phase topological defects arise, for example, in the form of magnetic edge dislocations\cite{uchida06, yu_10}. Analogous to edge dislocations in crystals and nematics, these magnetic edge dislocations naturally develop where helical spin textures of unequal phase meet, compensating for the local mismatch. At the bulk level, such line-like topological defects are often neglected as they affect only a small fraction of the volume. At the nanoscale, however, the defects and their dynamics become crucial as they can lead to significant perturbations in the electronic liquid in itinerant helimagnets. Thus, due to the close relation to the formation of the skyrmion phase and their general significance for the research on topological states, a detailed knowledge about the dynamics of topological defects in the helimagnetic state is highly desirable.

The probing of intrinsic micromagnetic instabilities at the nanoscale in a non-invasive way is a well-known challenge. Conventional microscopy methods, such as Lorentz transmission electron microscopy (TEM), magnetic force microscopy (MFM)\cite{hartmann99}, and scanning tunnelling microscopy (STM)\cite{Binnig86}, make either use of an electron beam or a magnetic probe tip and can themselves influence the behavior of the spin structure. As a consequence, it is difficult to unambiguously separate between intrinsic and extrinsic, probe-induced dynamical effects. Nitrogen-vacancy (NV) center-based magnetometry\cite{degen08,rondin14} is a new experimental method that, in principle, is capable of providing the desired information, but it has never been used to probe (helical) antiferromagnetic spin arrangements and rarely been applied under cryogenic conditions\cite{bouchard11,schafer14}.

In this work we study emergent micromagnetic dynamics in the helimagnetic phase of FeGe based on MFM, NV-center magnetometry, and Landau-Lifschitz-Gilbert (LLG) simulations. The MFM measurements reveal temperature-driven local changes in the magnetic domain structure, as well as jump-like collective movements of the helical spin texture that propagate over mesoscopic length scales. The collective movements are driven by the depinning and subsequent motion of topological magnetic edge dislocations by which the system relaxes its magnetic structure. Single-spin magnetometry experiments with NV centers, immobilized on the FeGe surface, show that these dynamics are intrinsic and highlight their stochastic nature. Coarse-grained LLG simulations are applied to analyze the microscopic magnetization dynamics. The simulations demonstrate that the movement and annihilation of topological defects plays a key role for the self-organization of the spin structure and the development of a long-range ordered helimagnetic ground state.

\begin{figure}[t]
\centering
\includegraphics[width=0.48\textwidth]{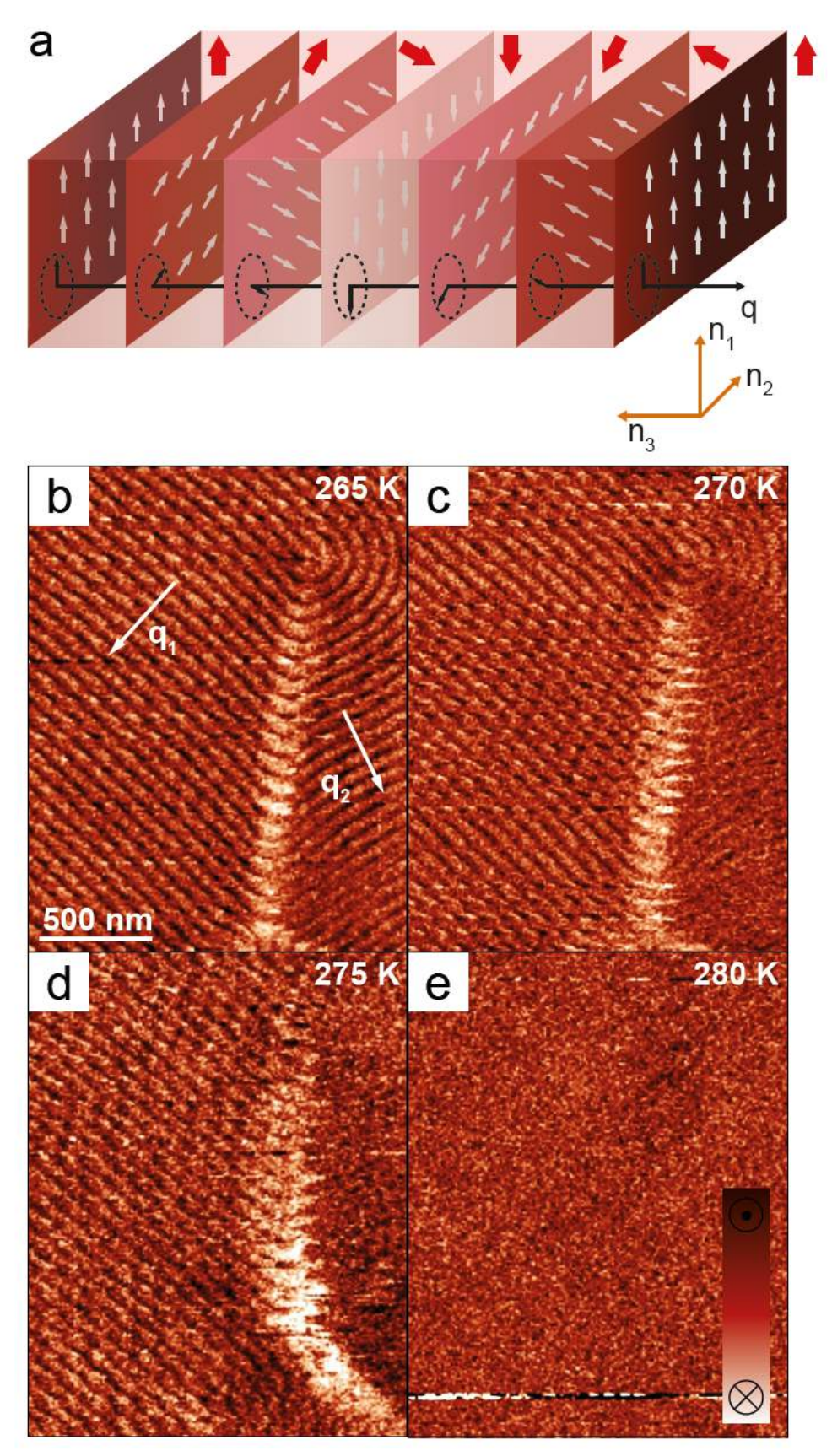}
\caption{\label{mfm1} \textbf{Temperature dependence of the helimagnetic domain structure in FeGe.}
\textbf{a}, Schematic of the helimagnetic spin order and wave vector $\mathbf q$. Color-coded planes indicate wave fronts of the magnetization ${\bf M}$ that are defined by uniformly oriented spins. \textbf{b}, MFM image in the helimagnetic state.  ${\bf q}_1$ and ${\bf q}_2$ indicate two magnetic domains.
\textbf{c} and \textbf{d}, With increasing temperature, local variations are observed in the the domain structure while the period of the helimagnetic order within domains is unaffected. \textbf{e}, The magnetic contrast vanishes when approaching $T_{\rm N}$.}
\end{figure}

For our studies on the dynamics of topological magnetic defects we choose cubic FeGe\cite{richardson67} as it exhibits helimagnetic order near room temperature with  $T_{\text N} \approx 280$~K and because its phase diagram is well-characterized\cite{wilhelm11}. The helical axis of the spin system is described by a wave vector {\bf q} which first points along the crystallographic $\left < 001 \right >$ direction (Fig.~\ref{mfm1}a), changing to the $\left < 111 \right >$ direction below 211~K upon cooling\cite{lebech89}. We begin our discussion with the spatially-resolved MFM measurements shown in Fig.~\ref{mfm1}b.  After cooling the sample to 265~K, alternating bright and dark lines are clearly visible, indicating a periodic magnetic structure. In order to relate the MFM data to the microscopic spin arrangement, we calculate the magnetic stray field for helimagnetic order with periodicity $\lambda$ and a constant magnetization amplitude $|{\bf M}|=M$ (see Fig.~\ref{mfm1}a for a schematic illustration of the helical spin structure). The magnetic structure can be described as
\begin{equation}\label{ansatz}
{\bf M}({\bf r})= M \{ {\bf n}_1 \cos( {\bf q}\cdot {\bf r}) + {\bf n}_2 \sin( {\bf q}\cdot{\bf r})\}  \; .
\end{equation}
Here, the ${\bf n}_i\ (i = 1, 2, 3)$ define a set of orthonormal unit vectors and ${\bf q} = {\bf n}_3 2 \pi/ \lambda $. For a sample with ${\bf q}$ lying in the surface plane and $z || {\bf n}_1$ being the probe distance above the sample surface, the spin helix described by Eq.~(\ref{ansatz}) leads to a magnetic stray field
\begin{equation}\label{Bbulk}
{\bf B}({\bf r})= \frac{M \mu_0}{2} \exp\left(-\frac{2\pi z}{\lambda}\right) \{ {\bf n}_1 \cos({\bf q}\cdot{\bf r}) + {\bf n}_3 \sin( {\bf q}\cdot {\bf r}) \} \; .
\end{equation}
Eq.~(\ref{Bbulk}) reflects that the periodicity of the stray field is equal to the periodicity $\lambda$ of the spin helix, and that the stray field exponentially drops with vertical decay length $\lambda/(2\pi)\approx 11$~nm.
Note that while the periodicities of ${\bf M}({\bf r})$  and ${\bf B}({\bf r})$ are the same, their rotation axes are orthogonal and defined by ${\bf n}_3$ and ${\bf n}_2$, respectively. The calculated magnetic stray field is in qualitative agreement with the MFM data in Fig.~\ref{mfm1} and we find $\lambda = 70\pm 5$~nm which is consistent with neutron scattering data\cite{lebech89}. The measurement in Fig.~\ref{mfm1}b further reveals micrometer-sized magnetic domains with different orientation of the wave vector {\bf q}, that are separated by a so-called vortex-free domain wall as detailed in Ref.~\onlinecite{li12}.

In order to investigate the stability of the helimagnetic order, we perform additional MFM scans at elevated temperature as illustrated in Fig.~\ref{mfm1}c-e. A comparison of the MFM images shows that the magnetic stray field associated with the structure of the domain wall between the ${\bf q}_1$- and ${\bf q}_2$-domain slightly varies with temperature, revealing a change in the length and orientation of the wall. The periodicity of the spin helix within the domains, by contrast, is robust against the temperature-driven variation in the domain pattern within the time-frame capture by the scan.

\begin{figure}
\centering
\includegraphics[width=0.48\textwidth]{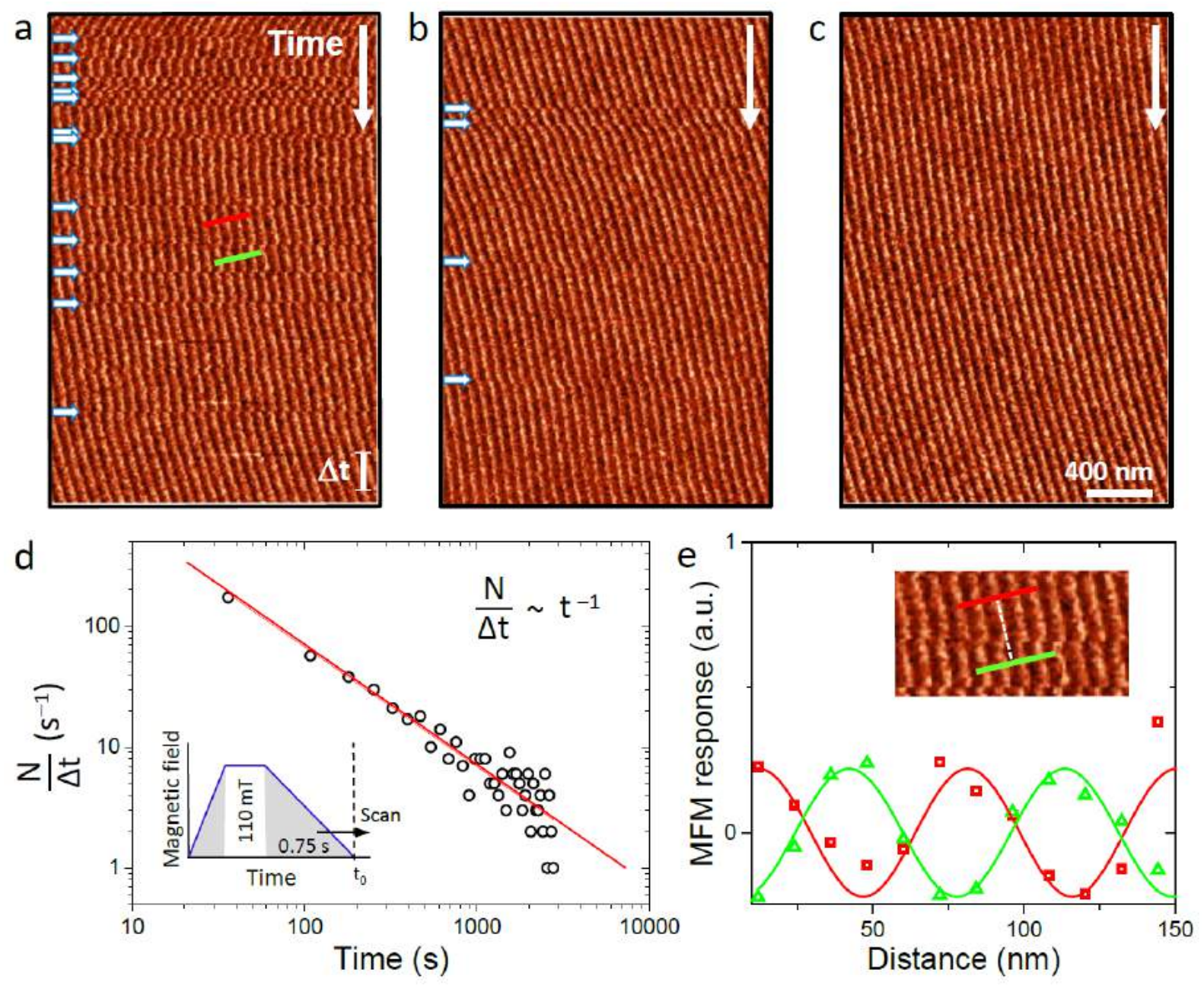}
\caption{\label{mfm2}\textbf{Dynamical phase jumps and relaxation behavior of helimagnetic order.}
 \textbf{a-c}, Representative MFM image series gained at the same sample posititon at $H = 0$~T after applying a magnetic field of 110~mT as sketched in the inset to \textbf{d}. Scan lines are recorded from top to bottom. The data reflect the emergence of stochastic collective jumps in the spin system, indicated by white arrows, which get less frequent as the scan progresses. The scale bar in \textbf{a} corresponds to a time frame of $\Delta t = 70$~s. \textbf{d}, Time-dependence of the rate of phase jumps, $N/\Delta t$, extracted from over 40 MFM image series as seen in \textbf{a-c} (see text for details). The graph reflects a relaxation that follows a power law with $ N/\Delta t\propto t^{-1}$. \textbf{e}, Evaluation of the change in period for the jump-like event shown in the inset (zoom-in to the area marked in \textbf{a}).}
\end{figure}

Occasionally, we observe jump-like collective movements in the helical spin structure while imaging. These jumps are especially visible after the spin system has been disturbed by a magnetic field or a change in temperature. Fig.~\ref{mfm2}a-c shows an MFM image series gained in the helical state after driving the system into the magnetic field-aligned  phase, as sketched in the inset of Fig.~\ref{mfm2}d. A systematic analysis of more than 40 time-dependent MFM experiments (performed at different sample positions) shows that the number of jumps $N$ per time interval $\Delta t = 70$~s follows a power law as known from slow relaxation processes (Fig.~\ref{mfm2}d)\cite{kubat}; we find $\frac{N}{\Delta t} \propto t^{-1}$. Interestingly, there is always a phase change associated with the individual jumps (Fig.~\ref{mfm2}e), often around 180$^{\circ}$.

\begin{figure}
\centering
\includegraphics[width=0.48\textwidth]{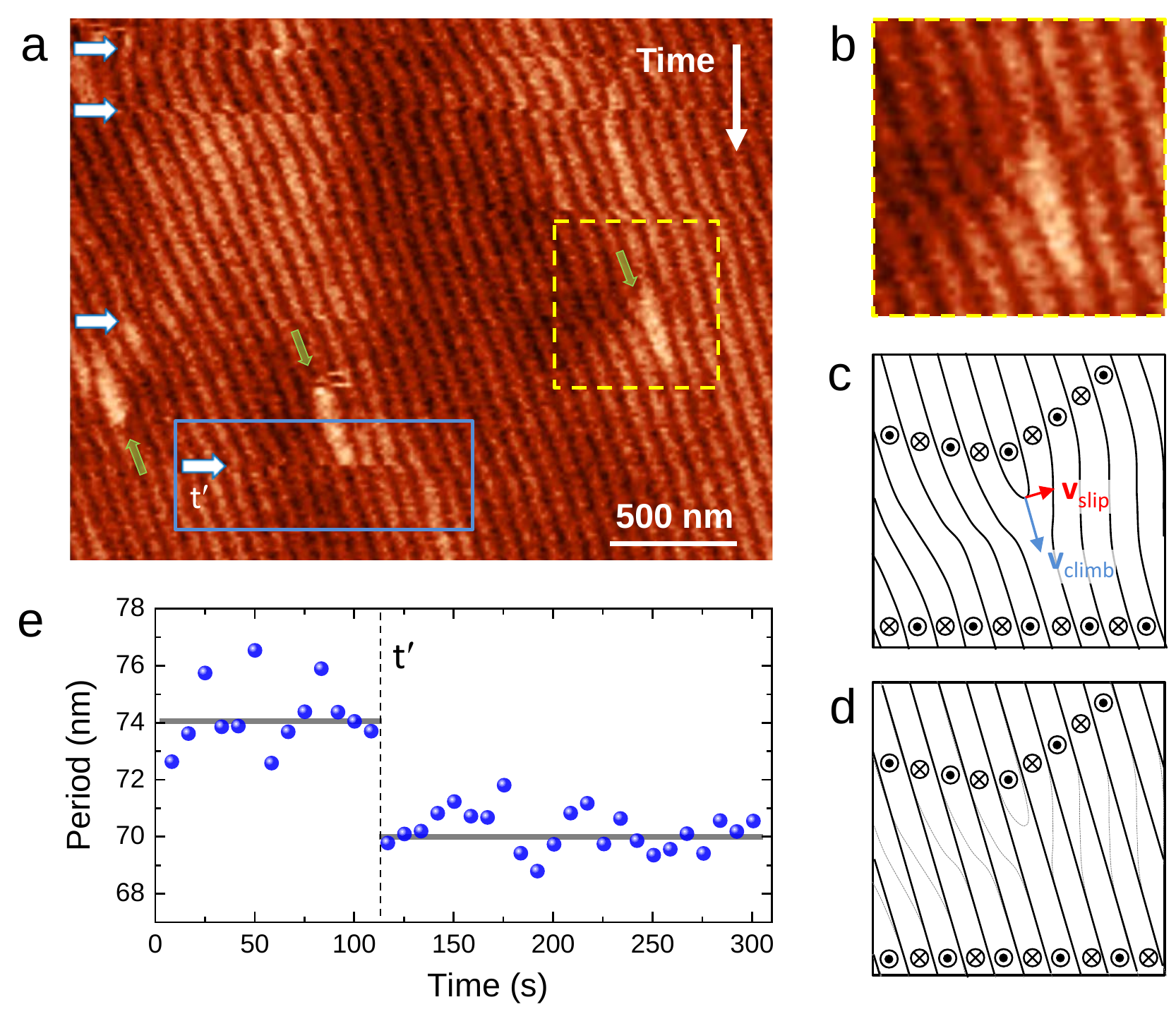}
\caption{\label{mfm3}\textbf{Static and mobile magnetic edge dislocations.}
\textbf{a}, MFM image of the helimagnetic structure at 266~K displaying several topological defects (magnetic edge dislocations; indicated by green arrows) and stochastic magnetization jumps (white arrows). Scan lines are recorded from top to bottom. \textbf{b}, Zoom-in to the area highlighted by the yellow dashed box in \textbf{a}, presenting a magnetic edge-dislocation. \textbf{c} Schematic illustration of the edge-dislocation seen in \textbf{b}. Black symbols indicate the direction of the out-of-plane component of the magnetic stray field, and arrows reflect the directions for slip and climb motions of the defect. \textbf{d} Illustration of the structure in \textbf{c} after the defect moved out of the field of view, yielding a 180° phase jump in the lower region of the sketch. \textbf{e} Evolution of the local magnetic period in the wake of the defect as function of time, evaluated for the blue solid box in \textbf{a}. At $t`$ the local mean period abruptly changes by about 4~nm (see white arrow in \textbf{a}).}
\end{figure}

Figure~\ref{mfm3} presents a possible relaxation mechanism, driven by the dynamics of magnetic defects, causing such collective jump-like movements.  The image in Fig.~\ref{mfm3}a is recorded at $T = 266$ K in a surface area with magnetic defects (marked by green arrows). These defects exhibit a locally enhanced magnetic stray field, leading to a brighter contrast level compared to the surrounding periodic spin structure. A closer inspection of the defects identifies them as magnetic edge dislocations as shown by the zoom-in and the corresponding sketch in Fig.~\ref{mfm3}b,c, respectively. Edge dislocation are line-like topological defects that, in the present case, allow the system to compensate for mismatches in the periodicity of its spin structure\cite{uchida06, hull11}.  The observation of magnetic edge dislocations in bulk FeGe complements earlier data obtained by Lorentz TEM on thin platelets of FeGe\cite{uchida08}, Fe$_{1-x}$Co$_x$Si\cite{uchida06, yu_10}, and BaFe$_{12-x-y}$Sc$_x$Mg$_y$O$_{19}$\cite{yu_12}.

The magnetization dynamics presented before in Fig.~\ref{mfm2} can be understood by assuming that magnetic edge dislocations spontaneously unpin and climb along the helical plane (that is, perpendicular to ${\bf q}$, see Fig.~\ref{mfm3}c,d). An example of such a spontaneous unpinning is shown in Fig.~\ref{mfm3}e.  Here, a jump of the spin system is captured at time $t^{\prime}$ that can be connected to the magnetic edge-dislocation which, at time $t<t^{\prime}$, was situated about $350$~nm above the solid blue box in Fig.~\ref{mfm3}c. The movement of this dislocation locally relaxes the initially stretched magnetic period ($\approx 74$~nm) to its equilibrium value of 70~nm, removing the tension that was associated with the previously pinned dislocation (see Fig.~\ref{mfm3}e).The relaxation of the tension is thus achieved by reducing the local density $n$ of edge dislocations so that we conclude $n \propto t^{-1}$.

The climbing of magnetic edge dislocations can also explain the tendency of the system to perform dynamical phase jumps of about 180$^\circ$ that can extend over many micrometers (see Fig.~\ref{mfm3}c,d for an illustration). The associated climb velocity is expected to be fast; a lower limit for $v_{\rm climb}$ can be derived from Fig.~\ref{mfm3}a based on the distance the defect travelled ($\Delta d\gtrsim 350$~nm) and the time difference between two consecutive scan lines ($\Delta t = 8$~s). We find $ v_{\rm{climb}} > 10^{-8}$~m/s, which would be comparable to slowly moving structural dislocations\cite{johnston59, edelin73}; however, our $v_{\rm{climb}}$ is a lower bound and the actual velocity may be much faster. In case of a defect-free magnetic environment we usually observe phase shifts to propagate across the entire field of view ($\gtrsim 10$~$\mu$m). Such a long-distance propagation is possible because of the incommensurability of the spin structure. Due to the incommensurability the free energy is independent of the helical phase and phase shifts cost no energy\cite{toledano87}. Thus, once launched, the energy gain associated with the local relaxation of the spin system can readily sustain the defect movement and the phase shift in its wake. Only the presence of pinned magnetic or structural defects, as well as domain walls, eventually halts the free propagation and confine the affected area.

\begin{figure*}
\centering
\includegraphics[width=0.8\textwidth]{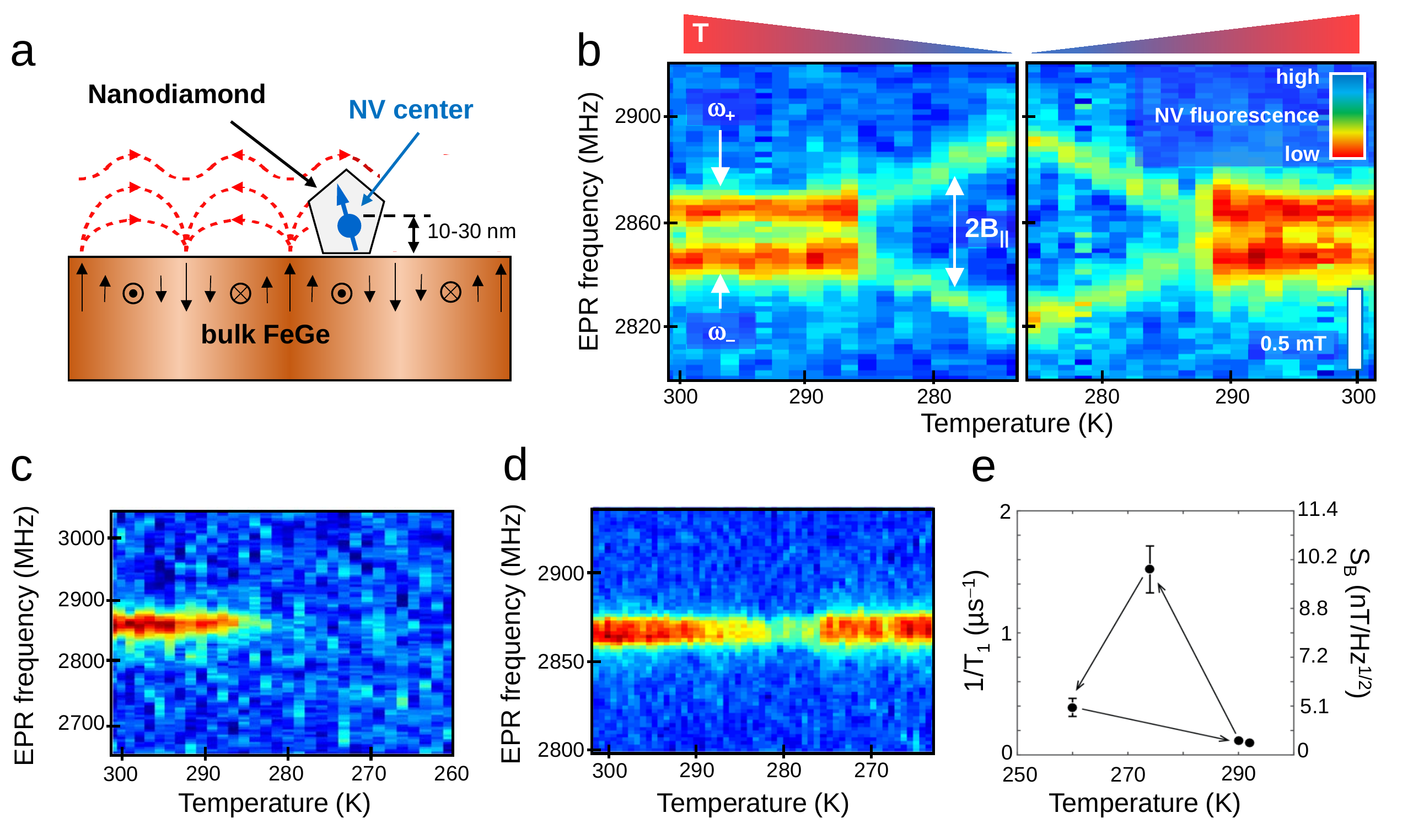}
\caption{\label{traces}\textbf{Dynamical variations in the local magnetic structure observed by single-spin magnetometry.}
\textbf{a}, Measurement schematic: A nanodiamond containing a single NV center is immobilized on the FeGe surface.
The local magnetic stray field (dashed red lines) induces Zeeman shifts to the NV center's  electronic spin transitions ($m_S=0\leftrightarrow m_S=\pm1$) that are measured using optically-detected EPR.  Black arrows indicate the helical spin texture of FeGe. \textbf{b}, Optically-detected EPR spectrum during cool-down (left panel) and warm-up (right panel), revealing the paramagnetic-to-helimagnetic phase transition at $T_{\rm N}\approx 286$~K.  Color coding reflects normalized fluorescence intensity. \textbf{c,d}, EPR signal of two NV centers, representing the end cases for high and low magnetic stray field strength, respectively. \textbf{e},  $T_1$ measured above, at, and below $T_{\rm N}$ for an NV center that showed a trace similar to \textbf{c}. Arrows indicate order of measurement. The scale on the right side provides the magnetic noise spectral density at the EPR frequency ($\sim 2.9\,{\rm GHz}$) calculated as $S_B = 2/(\gamma^2 T_1)$, where $\gamma$ is the electron gyromagnetic ratio.
}
\end{figure*}

In order to verify that the magnetization dynamics are intrinsic to FeGe and not triggered by the stray field of the MFM probe tip, we conduct a non-invasive magnetometry measurement with single NV centers in diamond\cite{degen08,rondin14,tetienne14,wolfe14} (Methods). As illustrated in Fig.~\ref{traces}a, we disperse diamond nanocrystals on the FeGe surface such that the NV centers are sufficiently close ($\sim 10-30$~nm) to pick up the local helimagnetic stray field.  Since the NV centers are immobilized on the surface, they can directly record any relative movement of the spin texture with respect to the underlying crystalline lattice of FeGe.
Magnetometry measurements are performed by monitoring the two electron paramagnetic resonance (EPR) transitions of the NV electronic spin using optical detection\cite{degen08}.  The difference between the two EPR frequencies, denoted by $\omega_{+}$ and $\omega_{-}$ in Fig.~\ref{traces}b, represents a Zeeman splitting that is proportional to the local magnetic stray field, 
\begin{equation}
B_{||} = \frac{1}{2\gamma}\sqrt{(\omega_{+}-\omega_{-})^2 - 4\delta^2} \; .
\end{equation}
Here, $\gamma = 2\pi \times 28$~GHz/T is the electron gyromagnetic ratio, $\delta$ is an additional splitting caused by strain in the nanocrystal, and $B_{||}$ is the component of ${\bf B}({\bf r})$ along the NV spin direction\cite{dolde11}.  At the same time, the sum of the two EPR frequencies can be used to monitor the local temperature $T$ via the (temperature-dependent) zero-field splitting parameter $D$,
\begin{equation}
D = \frac{1}{2}(\omega_{+}+\omega_{-}) \; ,
\end{equation}
with $D \approx 2867{\rm\,MHz} - 0.074{\rm\,MHz}\times(T-293{\rm\,K})/{\rm K}$\cite{acosta10}.  An EPR datapoint thus provides a simultaneous measurement of the local magnetic field and the local temperature.

Since the technique of NV magnetometry is relatively recent\cite{degen08} and has never been applied to the study of antiferromagnetic order, and rarely at low temperature\cite{bouchard11,schafer14}, we first demonstrate that the method is sensitive to the onset of helimagnetism. Fig.~\ref{traces}b presents a temperature scan across the phase transition.  Below $T_{\rm N} = 286 \pm 3$~K a pronounced Zeeman splitting is observed in the EPR signal, corresponding to an increase of $B_{||}$ from 0 to 1.2~mT.
The EPR splitting reversibly vanishes when returning to above $T_{\rm N}$.  The measurement thus clearly shows the sensitivity of NV magnetometry to the helimagnetic order.  The value of $T_{\rm N}$ found here is somewhat higher than expected from the MFM data (Fig.~\ref{mfm1}e) and literature values\cite{lebech89}, most likely due to the limited accuracy of the absolute temperature calibration of $D$.

Not all NV centers showed the response displayed in Fig.~\ref{traces}b, as the placement of nanodiamonds is stochastic and the vertical distance to the FeGe surface varies from NV center to NV center. Figure~\ref{traces}c,d show two end cases for a very close and a distant NV center. At close distance (Fig.~\ref{traces}c), the stray field is high and the EPR signal disappears entirely at $T<T_{\rm N}$ presumably due to fluorescence quenching\cite{epstein05}.  For a distant NV center (Fig.~\ref{traces}d), the stray field is too small to cause a measurable Zeeman splitting.  Interestingly, all traces show a pronounced reduction at $T_{\rm N}$ that is accompanied by a sharp reduction in the spin relaxation time $T_1$ (see Fig.~\ref{traces}e), indicating increased magnetic fluctuations at the phase transition.  Such fluctuations are expected from magnetic instabilities at the local scale that peak around $T_{\rm N}$\cite{janoschek2}.

\begin{figure}
\centering
\includegraphics[width=0.42\textwidth]{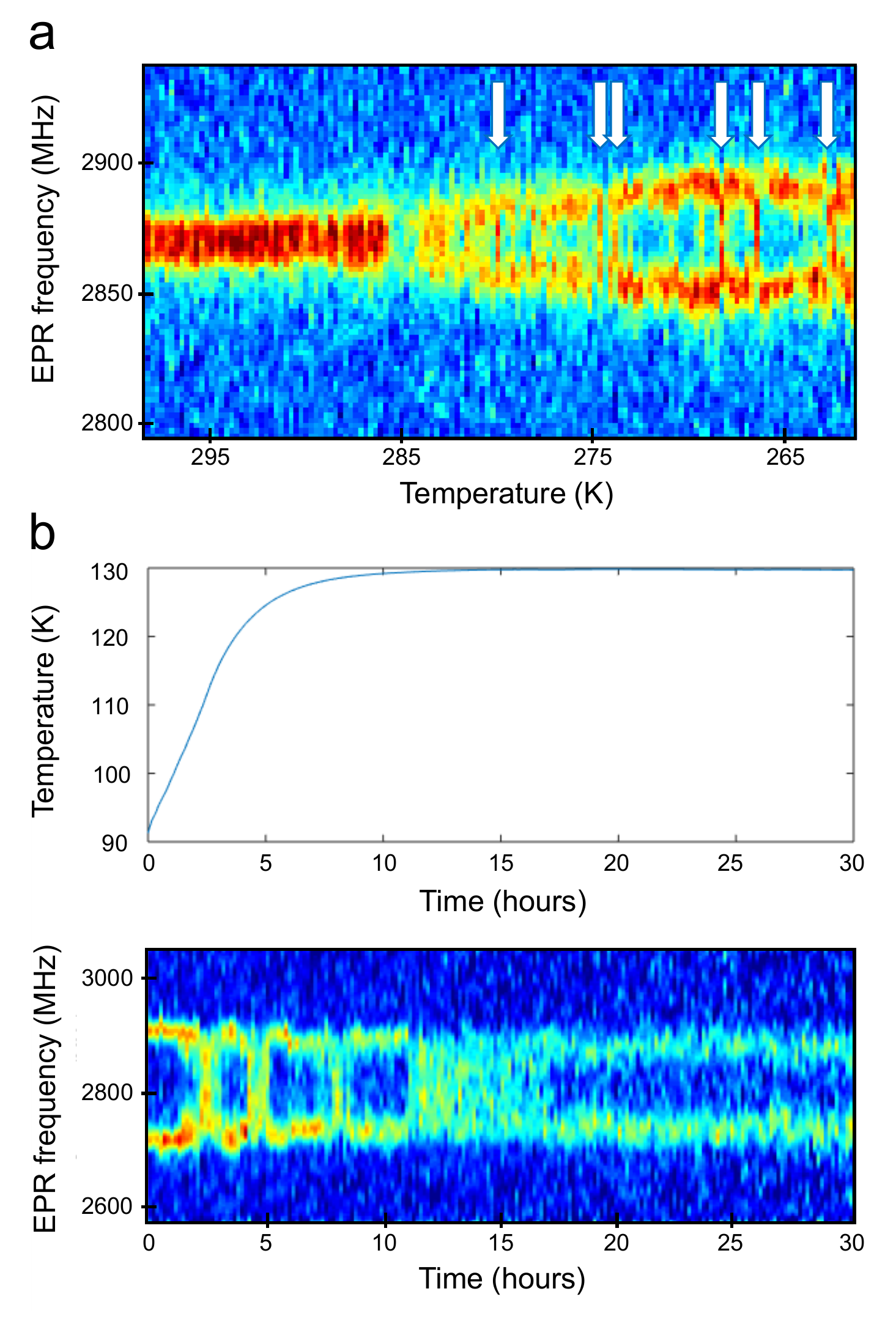}
\caption{\label{nv} \textbf{Transient breakdowns in the optically detected EPR signal.} \textbf{a}, EPR signal of a different NV center than in Fig ~\ref{traces}b showing again a splitting below $T_N$. The transient breakdowns in the EPR signal below $T_N$ indicate a sudden change of the local magnetization probed by the NV center (white arrows). \textbf{b}, Transient breakdowns in the EPR signal are detected as long as the temperature changes and vanish completely after the temperature is stabilized. 
}
\end{figure}

After discussing the NV center response to the onset of helimagnetism, we now turn to the detection of dynamical magnetic variations. Figure~\ref{nv}a presents the optically-detected EPR signal of a different NV center, recorded with decreasing temperature.  In agreement with Fig.~\ref{traces}b, a pronounced Zeeman splitting is visible below $T_{\rm N}$.  In addition we find that the splitting transiently breaks down while cooling as indicated by the white arrows in Fig.~\ref{nv}a.
Note that the maximum excursion of the Zeeman splitting varies only slowly with temperature, despite the many breakdowns, and assumes a roughly constant value below $T\sim 275$~K.  This behavior indicates that the breakdowns are associated with sudden changes in the orientation of the local magnetization, i.e., $|\textbf{B}|$ remains constant. These findings are consistent with the spin system's tendency towards phase jumps of about 180$^\circ$ obtained by MFM. Supplementary Fig.~1 shows that transients are also observed at temperatures far below $T_{\rm N}$. Opposite to the magnetically stimulated phase jumps observed by MFM (Fig.~\ref{mfm3}), however, the phase jumps observed with NV centers are caused by a change in temperature. The latter is reflected by Fig.~\ref{nv}b which confirms that the breakdowns are absent when the temperature is held constant for a long time.


In order to develop a microscopic model of the captured dynamics we perform simulations based on the LLG equation. We model the helimagnetism of FeGe with Heisenberg and DM exchange interactions obtained from electronic structure calculations as input parameters (Methods). This model yields a magnetic ground state with a perfect helical spin arrangement of period $\lambda\approx 100$~nm and $T_{\rm N} =240$~K, which is in fair agreement with the experimental observations. For $T > 0$~K magnetic fluctuations occur at the atomic scale and locally disturb the helimagnetic order (see Supplementary Fig.~2). Such magnetic excitations increase towards $T_{\rm N}$ and ultimately destroy the magnetic order, consistent with the MFM data shown in Fig.~\ref{mfm1} and the NV data in Fig.~\ref{traces}e.

\begin{figure}
\centering
\includegraphics[width=0.48\textwidth]{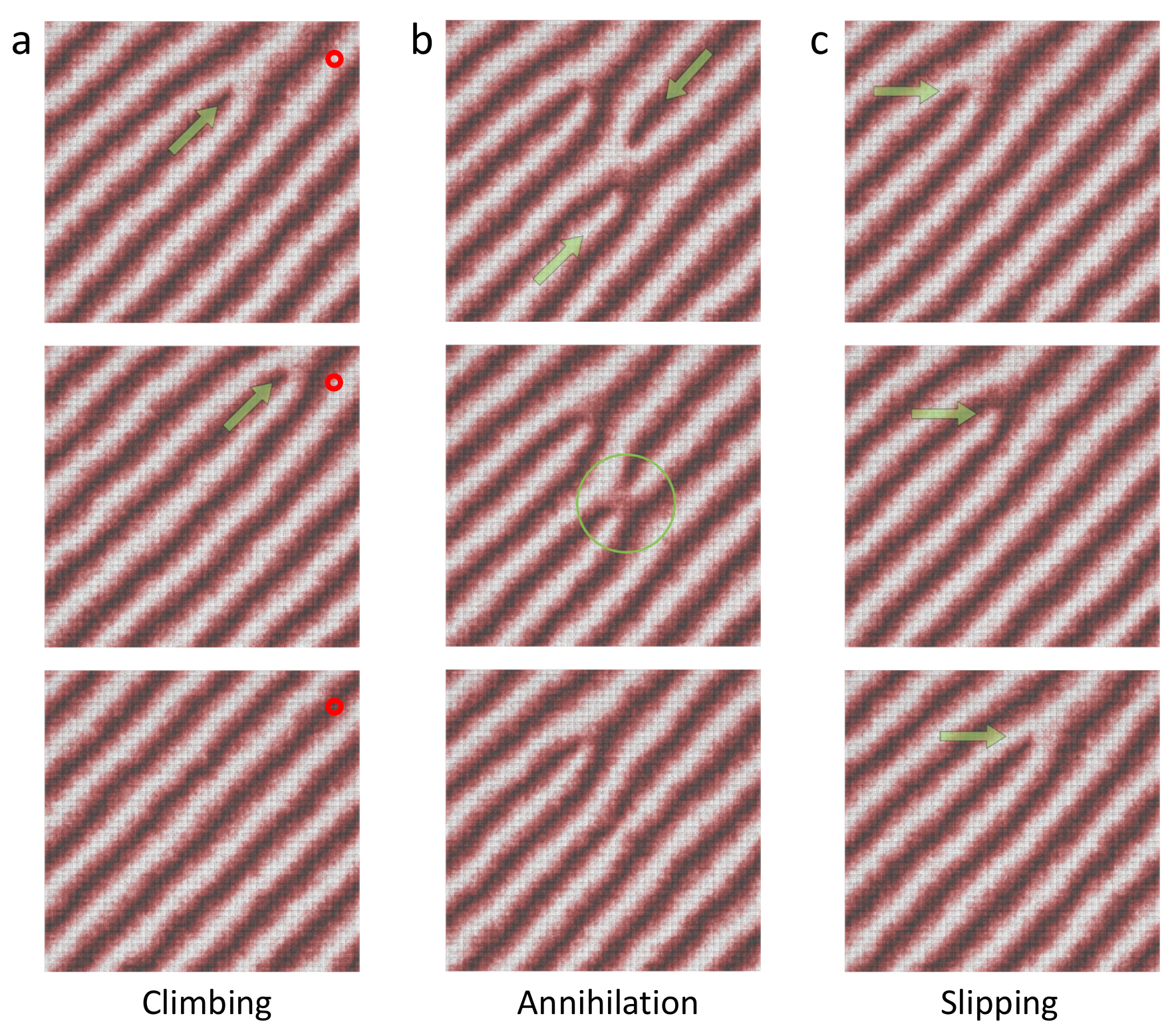}
\caption{\label{sim} \textbf{LLG simulations of emergent micromagnetic dynamics in the helimagntic state.} \textbf{a}, Simulations performed at $T = 0.5 \cdot T_N$ reveal that magnetic edge dislocations (green arrow) can climb through the helimagnetic structure. The movement locally relaxes the spin structure and induces a phase jump of 180° as seen, e.g., at the position marked by the red dot. \textbf{b,c}, Pair annihilation and slip motion of magnetic edge dislocations ($T = 0.5 \cdot T_N$).
}
\end{figure}

In addition to these local fluctuations, long-range magnetic excitations arise close to $T_{\rm N}$ that break the helimagnetic structure and naturally lead to the formation of positive and negative magnetic edge dislocations. Upon thermal quenching these edge dislocations remain quasi-stable as presented in Fig.~\ref{sim}. The quasi-stability is expected due to the topological nature of these magnetic excitations. The simulations further highlight that the magnetic edge dislocations are quite mobile. Figure~\ref{sim}a shows they can easily climb through the helical spin structure (${\bf v_{\rm{climb}}}\bot{\bf q}$). The climbing motion relaxes the local magnetic order and triggers phase shifts in the helimagnetic structure (see red circle in Fig.~\ref{sim}a), analogous to the illustration in Fig.~\ref{mfm3}c,d, which corroborates the above interpretation of our experimental data. Whenever positive and negative edge dislocations meet, they annihilate which further lowers the magnetic energy (Fig.~\ref{sim}b).

Interestingly, the micromagnetic simulations reveal that the magnetic edge dislocations can also move parallel to the wave vector ${\bf q}$, i.e., with ${\bf v_{\rm{slip}}}||{\bf q}$ as shown in Fig.~\ref{sim}c. The emergence of slip motions is surprising because slipping involves the destruction and creation of topolicial defects, but does not lead to an immediate relaxation of the spin system. Altogether, the micromagnetic calculations show that three types of magnetic defect dynamics, namely climbing, slipping, and pair annihilation, emerge at finite temperature in the helimagnetic phase. With this, the LLG simulations demonstrate a striking analogy between the dynamics of magnetic edge dislocations in FeGe and topological defects in crystals and nematics.
 
In summary, we have investigated the dynamics of topological magnetic defects in FeGe. By combining MFM, single-spin magnetometry with NV centers, and Landau-Lifschitz-Gilbert simulations we demonstrated that mobile magnetic edge dislocations play a key factor in the development of the helimagnetic ground state. Their movements help the system to order and reduce its free energy, but they also lead to stochastic perturbations that can propagate over microscopic distances and may explain the emergence of spontaneous magnetic instabilities in helical magnets\cite{ferriani08}. Such perturbations increase the noise level and need to be controlled adequately in envisaged device applications. We were able to generate micromagnetic instabilities both by a magnetic field ramp and small changes of temperature.  Analogous to magnetic monopoles\cite{milde13}, which are involved in the formation of skyrmion states, the magnetic edge dislocations discussed in our work are able to zip through the helimagnet. The obtained defect dynamics point towards fundamental similarities in the transportation of topological defects in electronic spin liquids and nematics, and reveal an intriguing connection between the micromagnetic dynamics in itinerant helimagnets and the self-organization of large-scale dynamic structures. 

\begin{flushleft}
\textbf{Acknowledgements} The authors thank M. Fiebig, Th. Lottermoser, and J. Rhensius for discussions and support. This work was supported by the Swiss National Science Foundation through grants 200021-149192, 200021-137520 and the NCCR QSIT, by the European Commission through project DIADEMS, and by the DARPA QuASAR programme. N.K. acknowledges funding through the JSPS Grant-in-Aids for Scientific Research (S) No. 24224009 and for Young Scientists (Start-up) No. 26886005 and A.B. through the Swedish Research Council (VR) and eSSENCE.\\

\textbf{Author contributions} A.D., K.C., and L.L. conducted the single-spin magnetometry and P.S. the MFM experiments.  K.K. and J.C. performed the LLG simulations supervised by A.B.; N.K. grew the FeGe single crystals under supervision of Y.T. M.G. introduced the idea of mobile magnetic edge dislocations.  D.M. devised and initiated the project.  A.D., P.S., C.D., and D.M. wrote the paper.  C.D. and D.M. supervised the work.  All authors discussed the results and contributed to their interpretation.

\end{flushleft}

\section*{Methods}
\textbf{Magnetic force microscopy.} For our experiments we prepared FeGe single crystals with a thickness of about 500~$\mu$m and a lateral extension of $1 \times 1$~mm. A surface roughness below 1~nm was achieved by chemo-mechanical polishing with silica slurry. All MFM data were recorded with a tip-surface distance of 30~nm using a home-built low-temperature holder based on a water-cooled three-stage peltier element which we implemented into a commercially available scanning probe microscope (NT-MDT).

\textbf{NV-center-based magnetometry.} Single-spin magnetometry experiments were carried out on a home-built confocal microscope housed in a dry optical cryostat (Montana Instruments Cryostation).  NV centers were illuminated using green 532-nm laser light and the fluorescence was detected through a 630-800-nm bandpass filter using a single photon counter module (Excelitas SPCM-AQRH).  Microwaves were generated using a synthesizer with adjustable frequency and power level (Quicksyn Phasematrix), amplified, and directed through a thin wire that passed in close proximity ($\sim 100\,\mu$m) of the NV center.  Optically-detected EPR spectra were taken by stepping the microwave frequency through resonance and recording the photon counts for each frequency. Nanodiamonds with a nominal diameter of 25 nm and typically $\sim 1$ NV per crystal (DiaScence, Van Moppes) were dispersed at low density on the FeGe surface such that single NV centers could be optically resolved.  The FeGe sample was mounted on a sapphire holder and thermally anchored on an OFHC copper sample stage that was cooled via a cold finger.  In order to avoid effects of local heating, the microwave wire was not allowed to touch the FeGe sample and low laser powers ($\approx 80\,\mu$W) were used for the optical readout.  The temperature of the sample was simultaneously monitored via the temperature-dependent EPR response of the NV center, and by a conventional thermometer attached to the sapphire holder.  We found that while local heating could be induced by high laser and microwave powers, it could be avoided by reducing the power level.

\textbf{Micromagnetic simulations.} We applied a multiscale approach to model the helimagnetism in FeGe. First, we obtained the electronic structure and magnetic properties by performing first principles calculations of FeGe in the B20 structure with a lattice parameter of 4.7Å\cite{lebech89}. The calculations were performed via the fully relativistic KKR method as implemented in the SPR-KKR package\cite{ebert1}. The shape of the potential was approximated via the Atomic Sphere Approximation (ASA) and the exchange correlation potential was treated via the Local Spin Density Approximation (LSDA) as parametrized by Vosko, Wilk, and Nusair (VWN)\cite{vosko}. Using the same method, both Heisenberg and DM exchange interactions were calculated\cite{ebert2}. These interactions define the spin Hamiltonian which served as the basis for the numerical simulations, where we used the Uppsala Atomistic Spin Dynamics package\cite{skubic} both for LLG and Monte Carlo simulations. The spin Hamiltonian was defined for atomic spins, but since the length scale for the helical spin state in FeGe is long compared to the atomic length scale, we performed coarse-grained simulations in addition to the atomistic simulations. In our coarse-graining scheme we still simulated discrete magnetic moments, but each discrete moment then represented the magnetization of a larger volume of the sample, from 1x1x1~nm to 5x5x5~nm. The interactions between the volume elements were then renormalized so that the effective exchange interactions corresponded to the same spin-wave stiffness and DM stiffness as in the atomistic situation. Coarse-graining the system like this gave a good description of long-wavelength fluctuations.


\newpage

\begin{figure*}
\centering
\includegraphics[width=0.85\textwidth]{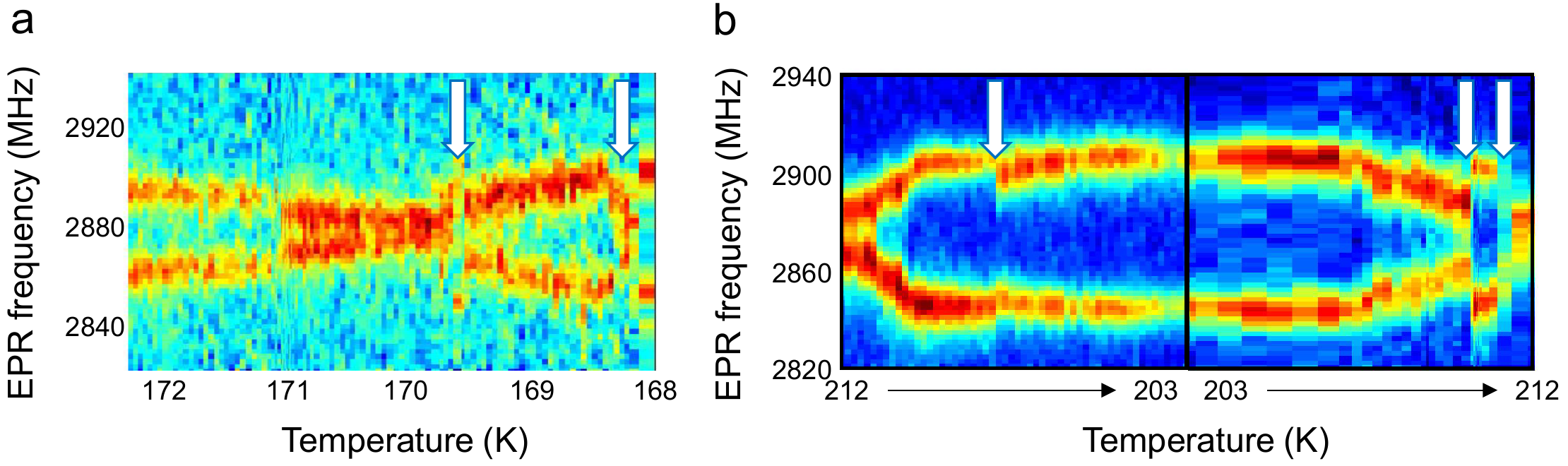}
\caption{\textbf{Supplementary Figure 1 - Quasi-reversible local dynamics probed by single-spin magnetometry.}
\textbf{a}, Besides the observed transient breakdowns in the EPR response (indicated by white arrows), a qualitatively different, second type of anomaly emerges in the response of specific NV centers as function of temperature. The Zeeman splitting continuously decreases with decreasing temperature between 172.5 K and 170 K and then reemerges with further decreasing temperature. The variation in the Zeeman splitting reflects a change in either the magnitude or orientation of the local magnetic stray field, with an amplitude of about 0.8 mT. Note that the gradual variation in the Zeeman splitting does not reflect a
magnetic phase transition. This is evident from the FeGe phase diagram and the fact that the temperature at which the EPR signal vanishes is specific to the evaluated NV center. As the temperature is lowered, the crystal lattice contracts, leading to slow changes in the local periodicity of the weakly pinned spin helix. \textbf{b}, The slow temperature-dependent Zeeman splitting is reversible during the reversible compression of the spin helix when the temperature is ramped between 213~K and 203~K and coexists with the sudden jumps discussed in the main text.}
\end{figure*}

\begin{figure*}
\centering
\includegraphics[width=0.6\textwidth]{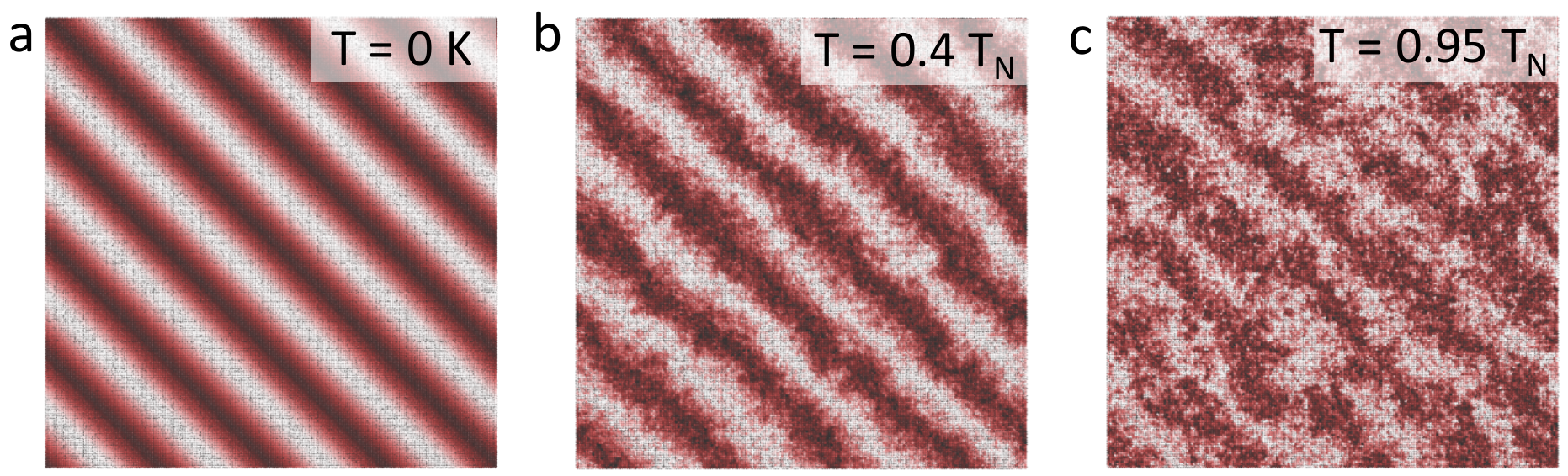}
\caption{\textbf{Supplementary Figure 2 - Simulated temperature-dependent helimagnetism in FeGe.}
\textbf{a}, In the magnetic ground state ($T=0$~K), FeGe displays a perfectly order helimagnetic spin arrangement, leading to a stripe-like patter when calculating the magnetic out-of-plane component (dark = $+M$, bright = $-M$). \textbf{b}, At finite temperature ($T = 0.4\cdot T_{\rm N}$) fluctuations arise at the atomic scale that locally perturb the helimagnetic order. \textbf{c}, Toward $T_{\rm N}$ ($T = 0.95\cdot T_{\rm N}$) the local fluctuations increase and eventually destroy the long-range magnetic order, leading to a paramagnetic state for $T > T_{\rm N}$.}
\end{figure*}

\end{document}